\documentclass{article}
\usepackage{spconf}
\usepackage{cite}
\usepackage{amsmath,amssymb,amsfonts,url,times,booktabs, tabularx,csquotes,hyperref}
\usepackage{algorithmic}
\usepackage{graphicx}
\usepackage{textcomp}
\usepackage{xcolor}
\usepackage{orcidlink}
\usepackage{adjustbox}
\usepackage{multirow,afterpage}
\usepackage{tikz,pgfplots}
\usepackage{amsbsy}  
\usepackage{acronym}
\pgfplotsset{compat=1.18}
\usetikzlibrary{shapes.geometric, arrows, calc, fit, shapes, backgrounds}

\newcommand{\BE}{\begin{equation*}\begin{aligned}}
\newcommand{\EE}{\end{aligned}\end{equation*}}

\acrodef{auc}[AUC-ROC]{area under the \acl{roc} curve}
\acrodef{asd}[ASD]{anomalous sound detection}
\acrodef{pauc}[pAUC]{partial area under the \acl{roc} curve}
\acrodef{ssl}[SSL]{self-supervised learning}
\acrodef{statex}[StatEx]{statistics exchange}
\acrodef{featex}[FeatEx]{feature exchange}
\acrodef{tmn}[TMN]{temporal mean normalization}
\acrodef{l3}[L3]{look, listen and learn}

\title{Self-Supervised Learning 
for Anomalous Sound Detection}

 \name{Kevin Wilkinghoff}
 \address{Fraunhofer FKIE, Fraunhoferstraße 20, 53343 Wachtberg, Germany\\
 kevin.wilkinghoff@ieee.org}
\begin{document}

\ninept
\maketitle
\begin{sloppy}

\begin{abstract}
State-of-the-art \ac{asd} systems are often trained by using an auxiliary classification task to learn an embedding space.
Doing so enables the system to learn embeddings that are robust to noise and are ignoring non-target sound events but requires manually annotated meta information to be used as class labels.
However, the less difficult the classification task becomes, the less informative are the embeddings and the worse is the resulting \ac{asd} performance.
A solution to this problem is to utilize \ac{ssl}.
In this work, \ac{featex}, a simple yet effective \ac{ssl} approach for \ac{asd}, is proposed.
In addition, \ac{featex} is compared to and combined with existing \ac{ssl} approaches.
As the main result, a new state-of-the-art performance for the DCASE2023 \ac{asd} dataset is obtained that outperforms all other published results on this dataset by a large margin.
\end{abstract}


\acresetall
\section{Introduction}
\label{sec:intro}
In contrast to supervised learning, \ac{ssl} \cite{liu2022audio} does not require manually annotated class labels.
Instead, data is augmented in different, specifically chosen ways, each defining another artificially created class and a model is taught to discriminate between these classes.
The underlying assumption is that the model needs to understand the structure of the data to correctly predict the artificially introduced classes.
\ac{ssl} is a type of unsupervised learning and has been successfully applied to learning speech representations \cite{mohamed2022self} or general purpose audio representations \cite{niizumi2021byol,niizumi2023byol} using large datasets of unlabeled data.
\par
\ac{ssl} has also been applied to \ac{asd}:
In \cite{inoue2020detection}, combinations of pitch-shifting and time-stretching are used to create additional classes.
\cite{lopez2020speaker} uses linear combinations of similar target sounds, created by applying mixup \cite{zhang2017mixup}, as pseudo-anomalous classes.
\Ac{statex} \cite{chen2023effective} is an \ac{ssl} approach that mixes first- and second-order statistics of time-frequency representations to create new classes.
In \cite{nejjar2022dg-mix}, a modified version of variance-invariance-covariance regularization \cite{bardes2022vicreg}, called domain generalization mixup, is used to pre-train autoencoders.
Note that some works on \ac{asd} use the term \ac{ssl} for supervised learning of embeddings with auxiliary classification tasks \cite{giri2020self,dohi2021flow}.
Since classification tasks require manually labeled data whereas \ac{ssl} does not require any manual annotations, we will use two different terms.
\par
In \cite{wilkinghoff2024why}, it was shown that class labels alone are very beneficial to detect anomalous sounds in noisy conditions as the model learns to closely monitor the target sounds.
When not using class labels, many non-target sounds contained in an acoustic scene have a much stronger impact than subtle changes of the target sounds that need to be detected in order to identify anomalous sounds.
However, as the available classes become less similar to each other, the embeddings acquired through the discrimination of these classes capture less meaningful information.
In these cases, also utilizing \ac{ssl} is a very promising approach to increase the degree of information being captured and thus is expected to improve the \ac{asd} performance. 
\par
The goal of this work is to investigate different approaches of using \ac{ssl} for \ac{asd}.
The contributions are the following:
First, two existing \ac{ssl} approaches for \ac{asd}, namely mixup and \ac{statex}, are reviewed.
Second, \ac{featex}, a novel \ac{ssl} approach for \ac{asd}, and a combination with the other two \ac{ssl} approaches are proposed.
In experimental evaluations conducted on the DCASE2022 \ac{asd} dataset and the DCASE2023 \ac{asd} dataset, it is shown that the proposed approach improves the performance over a baseline system not using any \ac{ssl}.
As a result, a new state-of-the-art performance significantly outperforming all published \ac{asd} results is obtained on the DCASE2023 \ac{asd} dataset\footnote{An open-source implementation of the system is available at: \url{https://github.com/wilkinghoff/ssl4asd}}.

\section{State-of-the-art Baseline system}
\label{subsec:baseline}

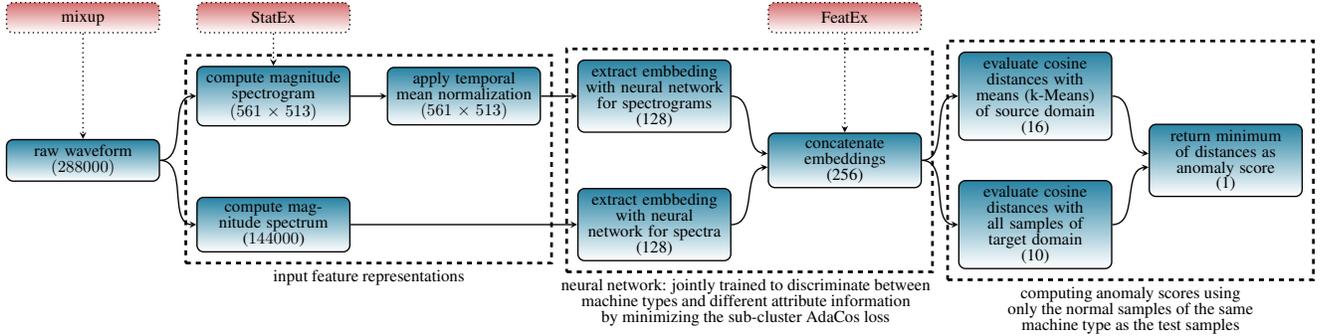
\begin{figure*}[t]
	\centering
	\begin{adjustbox}{max width=\textwidth}
		\usetikzlibrary{shapes.geometric, arrows, calc, fit}
\tikzset{box/.style={top color=teal!85!blue!85,shade,draw, rectangle, rounded corners, thick, node distance=1cm, minimum width=2.3cm, text centered, minimum height=2em, text width=3cm}}
\tikzset{box_red/.style={top color=teal!25!red!65,dotted,shade,draw, rectangle, rounded corners, thick, node distance=1cm, minimum width=2.3cm, text centered, minimum height=2em, text width=3cm}}
\tikzset{container/.style={draw, rectangle, dashed, ultra thick, inner sep=0.7em,minimum width=2cm}}
\tikzset{container2/.style={,draw, rectangle, dashed, ultra thick, inner sep=0.7em,minimum width=2cm}}
\tikzset{font={\fontsize{10pt}{10}\selectfont}}
\tikzstyle{arrow} = [thick, ->, >=stealth]
\begin{tikzpicture}
\node(wav)[box,node distance=1.65cm]{raw waveform\\ $(288000)$};
\node(mixup)[box_red,above of=wav, node distance=3cm]{mixup};
\node(log_mel)[box,right of=wav, yshift=1.35cm, node distance=4cm]{compute magnitude spectrogram $(561\times513)$};
\node(cmn)[box,right of=log_mel, node distance=4cm]{apply temporal mean normalization $(561\times513)$};
\node(fft)[box,right of=wav, yshift=-1.35cm, node distance=4cm]{compute magnitude spectrum\\ ($144000$)};
\node(nnmel)[box,right of=cmn, node distance=4cm]{extract embbeding with neural network for spectrograms\\(128)};
\node(statex)[box_red,above of=log_mel, node distance=1.65cm]{StatEx};
\node(nnfft)[box,right of=fft, node distance=8cm]{extract embbeding with neural network for spectra\\(128)};
\node(concat)[box,right of=nnfft, yshift=1.35cm, node distance=4cm]{concatenate embeddings\\ (256)};
\node(featex)[box_red,above of=concat, node distance=3cm]{FeatEx};
\node(gmmconcat1)[box,right of=concat, node distance=4cm, yshift=-1.35cm]{evaluate cosine distances with all samples of target domain\\(10)};
\node(gmmconcat2)[box,right of=concat, node distance=4cm, yshift=1.35cm]{evaluate cosine distances with means (k-Means) of source domain\\(16)};
\node(llh)[box,right of=gmmconcat2, node distance=4cm, yshift=-1.35cm]{return minimum of distances as anomaly score\\ (1)};
\begin{scope}[on background layer]
\node[container2, fit=(nnfft)(nnmel)(concat)] (alltrain) {};
\node at (alltrain.south west) [text centered,below right,node distance=0 and 0, align=center,xshift=-0.2cm] (alltraintxt) {neural network: jointly trained to discriminate between\\ machine types and different attribute information\\by minimizing the sub-cluster AdaCos loss};
\node[container2, fit=(fft)(log_mel)(cmn)] (frontend) {};
\node at (frontend.south) [text centered,below,node distance=0 and 0, align=center,xshift=0cm] (frontendtxt) {input feature representations};
\node[container2, fit=(gmmconcat1)(gmmconcat2)(llh)] (backend) {};
\node at (backend.south) [text centered,below,node distance=0 and 0, align=center,xshift=0cm] (backendtxt) {computing anomaly scores using\\only the normal samples of the same\\ machine type as the test samples};
\end{scope}

\draw[arrow](wav.0) [out=0, in=180] to (log_mel.180);
\draw[arrow](log_mel.0) [out=0, in=180] to (cmn.180);
\draw[arrow](cmn.0) [out=0, in=180] to (nnmel.180);
\draw[arrow](wav.0) [out=0, in=180] to (fft.180);
\draw[arrow](fft.0) [out=0, in=180] to (nnfft.180);
\draw[arrow](nnfft.0) [out=0, in=185] to (concat.185);
\draw[arrow](nnmel.0) [out=0, in=175] to (concat.175);
\draw[arrow](concat.0) [out=0, in=180] to (gmmconcat1.180);
\draw[arrow](concat.0) [out=0, in=180] to (gmmconcat2.180);
\draw[arrow](gmmconcat1.0) [out=0, in=185] to (llh.185);
\draw[arrow](gmmconcat2.0) [out=0, in=175] to (llh.175);

\draw[arrow,dotted](mixup.270) [out=270, in=90] to (wav.90);
\draw[arrow,dotted](statex.270) [out=270, in=90] to (log_mel.90);
\draw[arrow,dotted](featex.270) [out=270, in=90] to (concat.90);

\end{tikzpicture}
	\end{adjustbox}
	\caption{Structure of the baseline system (blue boxes). \acs{ssl} approaches are colored in red. Representation size in each step is given in brackets. This figure is adapted from  \cite{wilkinghoff2023fkie} and originally adapted from \cite{wilkinghoff2023design}.}
	\label{fig:system_structure}
\end{figure*}
Throughout this work, the \ac{asd} system presented in our prior work \cite{wilkinghoff2023design}, which is trained in a supervised manner by using an auxiliary classification task, is used as a baseline system.
Our goal is to improve the performance obtained with this system by applying \ac{ssl} approaches.
This system ranked $4$th in the DCASE2023 Challenge \cite{wilkinghoff2023fkie} and the winning team of the challenge \cite{junjie2023anomalous} extended this system by adding an attention mechanism to the embedding model.
This shows that the \ac{asd} system can be considered state-of-the-art, which justifies choosing it as a baseline system.
\par
An overview of the baseline system can be found in \autoref{fig:system_structure}.
The system is based on learning discriminative embeddings by using all available meta information, in this case all combinations of machine types, machine IDs and attribute information, as classes.
To this end, two convolutional sub-networks are used and their outputs are concatenated to obtain a single embedding.
One sub-network utilizes the full magnitude frequency spectrum as an input representation to ensure the highest possible frequency resolution.
The other sub-network uses magnitude spectrograms while subtracting the temporal mean to remove static frequency information and make the input representation more different from the other one.
The neural network is trained for $10$ epochs using a batch size of $64$ by minimizing the angular margin loss sub-cluster AdaCos \cite{wilkinghoff2021sub} with $16$ sub-clusters.
To improve the resulting \ac{asd} performance, no bias terms are used in any layer of the networks and the cluster centers are randomly initialized and not adapted during training. 
Apart from mixup, no other data augmentation technique is used.
As a backend, k-means is applied to the embeddings obtained with the normal training samples of the source domain, for which many training samples are available.
The smallest cosine distance to these means and all samples of the target domain, which differs from the target domain and for which only very few training samples are available, is used as an anomaly score. 
More details about this baseline system can be found in \cite{wilkinghoff2023design}.

\section{Self-Supervised Learning Approaches}
\label{sec:method}

In this section, two \ac{ssl} approaches for training \ac{asd} systems are reviewed, namely mixup \cite{zhang2017mixup} and \ac{statex} \cite{chen2023effective}. Furthermore, a third approach, called \ac{featex}, is proposed and described in detail.
Last but not least, a combined \ac{ssl} approach, which jointly uses all three discussed approaches, is presented.

\subsection{Mixup}
\label{subsec:mixup}
The data augmentation technique mixup \cite{zhang2017mixup}, which uses linear interpolations between two training samples and their corresponding categorical labels, is widely applied for \ac{asd} \cite{junjie2023anomalous,wilkinghoff2023fkie,yafei2023unsupervised,wang2023first,JiaJun2023self}.
When applying mixup, two randomly chosen training samples $x_1,x_2$ and their corresponding categorical class labels $y_1,y_2\in[0,1]^{N_\text{classes}}$ with $N_\text{classes}\in\mathbb{N}$ denoting the number of classes are combined by setting
\BE
x_\text{new}&=\lambda x_1&&+(1-\lambda)x_2\\
y_\text{new}&=\lambda y_1&&+(1-\lambda)y_2
\EE
using a random mixing coefficient $\lambda\in[0,1]$.
Although mixup is just a data augmentation technique that does not introduce new classes, it can also be seen as a form of \ac{ssl} that requires class labels because the supervised training objective is essentially extended to also predicting the mixing coefficient in addition to the original classes.
In \cite{chen2023effective,wilkinghoff2023fkie}, mixup was used to create additional pseudo-anomalous classes by treating mixed-up samples as belonging to other classes as non-mixed samples, similar to the approach proposed in \cite{inoue2020detection}.
In our experiments, this approach did not improve the \ac{asd} performance over applying mixup regularly but even degraded the performance for some machine types.
It is also possible to only predict the mixing coefficient by ignoring the class labels, making mixup a purely self-supervised approach.
However, for noisy audio data it is highly beneficial to utilize all available meta information for classification in order to teach the system to closely monitor the machine sounds of interest and ignore the background noise and other non-target events \cite{wilkinghoff2024why}.
Throughout this work, we used a probability of $100\%$ for applying mixup when training the baseline system, and a probability of $50\%$ when also applying any other \ac{ssl} approach. 
\begin{table*}[t]
	\centering
	\caption{Harmonic means of AUCs and pAUCs taken over all machine IDs obtained when using different \ac{ssl} approaches. Highest AUCs and pAUCs in each row are highlighted in bold letters. Arithmetic mean and standard deviation over five independent trials are shown.}
\begin{adjustbox}{max width=\textwidth}
	\begin{tabular}{lll|ll|ll|ll|ll|ll|ll}
		\toprule
		\multicolumn{3}{c}{}&\multicolumn{2}{c}{baseline \cite{wilkinghoff2023design}}&\multicolumn{2}{c}{\acs{statex} \cite{chen2023effective} variant}&\multicolumn{2}{c}{\acs{featex}}&\multicolumn{2}{c}{regular and \acs{statex} \cite{chen2023effective} variant}&\multicolumn{2}{c}{regular and \acs{featex}}&\multicolumn{2}{c}{proposed approach}\\
		dataset&split&domain&AUC&pAUC&AUC&pAUC&AUC&pAUC&AUC&pAUC&AUC&pAUC&AUC&pAUC\\
		\midrule
        \midrule
		DCASE2022&dev&source&$84.2\pm0.8$\%&$76.5\pm0.9$\%&$80.5\pm1.8$\%&$69.5\pm1.9$\%&$82.1\pm0.9$\%&$72.8\pm0.8$\%&$85.2\pm0.9$\%&$77.5\pm1.2$\%&$85.1\pm0.9$\%&$76.3\pm1.8$\%&\pmb{$86.0\pm0.9$\%}&\pmb{$77.6\pm0.8$\%}\\
		DCASE2022&dev&target&$78.5\pm0.9$\%&$62.5\pm0.9$\%&$75.3\pm1.7$\%&$60.0\pm0.9$\%&$77.2\pm1.0$\%&$62.8\pm0.6$\%&\pmb{$78.9\pm0.9$\%}&$63.2\pm1.6$\%&$77.9\pm0.9$\%&$62.7\pm0.8$\%&$78.2\pm0.7$\%&\pmb{$64.4\pm1.1$\%}\\
		DCASE2022&dev&mixed&$81.4\pm0.7$\%&$66.6\pm0.9$\%&$76.4\pm1.5$\%&$62.4\pm1.2$\%&$78.5\pm0.6$\%&$65.2\pm0.6$\%&$82.2\pm0.6$\%&$67.0\pm1.0$\%&$81.6\pm0.7$\%&$67.0\pm0.9$\%&\pmb{$82.5\pm0.8$\%}&\pmb{$68.2\pm1.1$\%}\\
		\midrule
		DCASE2022&eval&source&$76.8\pm0.8$\%&$65.8\pm0.2$\%&$74.2\pm0.6$\%&$61.6\pm1.4$\%&$76.3\pm0.9$\%&$64.5\pm1.2$\%&$76.9\pm0.4$\%&$65.8\pm0.9$\%&\pmb{$78.1\pm0.4$\%}&\pmb{$67.0\pm1.1$\%}&$77.7\pm0.8$\%&\pmb{$67.0\pm0.5$\%}\\
		DCASE2022&eval&target&$69.8\pm0.5$\%&$59.7\pm1.1$\%&$70.6\pm0.5$\%&$59.0\pm0.7$\%&\pmb{$72.3\pm0.6$\%}&$61.0\pm0.7$\%&$71.2\pm0.3$\%&$60.3\pm0.7$\%&$72.2\pm0.4$\%&\pmb{$61.3\pm0.5$\%}&$71.6\pm1.0$\%&$61.2\pm0.9$\%\\
		DCASE2022&eval&mixed&$73.4\pm0.5$\%&$59.8\pm0.8$\%&$72.2\pm0.3$\%&$58.2\pm0.7$\%&$73.9\pm0.5$\%&$60.0\pm0.9$\%&$73.9\pm0.3$\%&$59.9\pm0.6$\%&\pmb{$74.9\pm0.4$\%}&\pmb{$61.5\pm0.6$\%}&$74.2\pm0.3$\%&$61.2\pm0.3$\%\\
		\midrule
        \midrule
		DCASE2023&dev&source&$69.8\pm1.8$\%&$60.9\pm0.9$\%&$67.8\pm1.5$\%&$59.2\pm0.8$\%&$68.4\pm1.0$\%&$60.2\pm0.5$\%&$70.3\pm1.8$\%&$62.0\pm1.6$\%&\pmb{$72.9\pm2.0$\%}&\pmb{$63.0\pm1.3$\%}&$71.2\pm1.6$\%&$62.7\pm1.3$\%\\
		DCASE2023&dev&target&$72.3\pm1.8$\%&$55.6\pm0.9$\%&$69.7\pm1.7$\%&$54.7\pm1.1$\%&$74.4\pm0.7$\%&\pmb{$57.6\pm1.0$\%}&$72.2\pm1.4$\%&$56.2\pm1.2$\%&\pmb{$75.7\pm0.8$\%}&$57.0\pm1.6$\%&$75.0\pm1.5$\%&$56.1\pm1.4$\%\\
		DCASE2023&dev&mixed&$71.3\pm0.6$\%&$56.1\pm0.8$\%&$69.0\pm1.2$\%&$55.7\pm1.0$\%&$71.7\pm0.4$\%&$57.5\pm0.7$\%&$71.2\pm0.7$\%&$57.0\pm1.4$\%&\pmb{$74.4\pm1.0$\%}&\pmb{$58.0\pm1.4$\%}&$73.1\pm0.9$\%&$57.3\pm0.6$\%\\
		\midrule
		DCASE2023&eval&source&$72.5\pm0.8$\%&$62.4\pm1.2$\%&$70.0\pm1.2$\%&$59.7\pm0.9$\%&$69.3\pm2.0$\%&$59.3\pm1.2$\%&$72.4\pm2.4$\%&$62.4\pm1.3$\%&\pmb{$75.9\pm1.0$\%}&$62.9\pm1.3$\%&$75.5\pm0.8$\%&\pmb{$64.5\pm0.6$\%}\\
		DCASE2023&eval&target&$63.1\pm2.6$\%&$57.5\pm0.8$\%&$66.7\pm1.8$\%&$58.4\pm0.7$\%&\pmb{$69.1\pm1.3$\%}&$59.0\pm1.3$\%&$65.9\pm1.9$\%&$59.0\pm1.4$\%&$66.5\pm1.9$\%&$58.3\pm1.0$\%&$68.7\pm2.2$\%&\pmb{$59.3\pm0.7$\%}\\
		DCASE2023&eval&mixed&$67.9\pm1.0$\%&$58.8\pm0.8$\%&$65.8\pm0.6$\%&$57.1\pm0.8$\%&$68.1\pm1.2$\%&$58.1\pm0.9$\%&$69.5\pm1.8$\%&$60.7\pm0.9$\%&$71.1\pm1.1$\%&$60.1\pm1.3$\%&\pmb{$72.6\pm0.7$\%}&\pmb{$61.6\pm0.5$\%}\\
        \midrule
		\bottomrule
	\end{tabular}
\end{adjustbox}
\label{tab:results}
\end{table*}
\subsection{Statistics exchange}
\label{subsec:statex}
The idea of \ac{statex} \cite{chen2023effective} is to artificially create new classes of pseudo anomalies by exchanging the first- and second-order statistics of the time-frequency representations of two training samples $x_1,x_2$ along the temporal or frequency dimension.
Mathematically, this corresponds to generating a new sample $x_\text{new}\in\mathbb{R}^{T\times F}$ by setting
\BE x_\text{new} = \frac{x_1-\mu_1}{\sigma_1}\sigma_2+\mu_2\EE
where $x_1,x_2\in\mathbb{R}^{T\times F}$ denote the time-frequency representations of two random training samples.
$\mu_1,\mu_2$ denote the first-order statistics of these samples along the time or frequency dimension, and $\sigma_1,\sigma_2$ denote the second-order statistics along the same dimension.
Each possible combination of classes defines a new class, increasing the original number of classes $N_\text{classes}\in\mathbb{N}$ by a quadratic term $N_\text{classes}^2$.
\par
In this work, we used a variant of \ac{statex} with the following modifications:
For the sake of simplicity, we always use the complete frequency band and all time steps to calculate the statistics whereas in the original definition subbands are used \cite{chen2023effective}.
Third, we teach the model to predict the class of the original sample $x_1$ and the class of the other sample $x_2$, whose statistics are used.
For categorical class labels $y_1,y_2$, we do this by concatenating the labels:
\BE y_\text{new}=(\textbf{0},0.5\cdot y_1,0.5\cdot y_2)\in[0,1]^{3N_\text{classes}}\EE
where $\textbf{0}=(0,...,0)\in[0,1]^{N_\text{classes}}$.
Hence, the number of classes is only tripled and thus the number of parameters, which, due to the cluster centers, proportionally increases with the number of classes, does not explode.
Furthermore, this enables a simple combination with other data augmentation techniques, for which more than a single class are assigned to each sample, such as mixup.
In \cite{wilkinghoff2023design}, it has been shown that removing the temporal mean from the spectrograms improves the resulting \ac{asd} performance.
Thus, in the variant used in this paper, we applied \ac{statex} to the frequency axis, i.e. we only used temporal \ac{statex}.
Furthermore, two feature branches are used in the baseline system.
Hence, only temporal \ac{statex} has been applied to the spectrogram representations.
Throughout this work, we used a probability of $50\%$ for applying \ac{statex} during training and also applied mixup.
In case \ac{statex} is not applied, the new label of training sample ${x_\text{new}=x_1}$ is set to
\BE y_\text{new}=(y_1,\textbf{0},\textbf{0})\in[0,1]^{3N_\text{classes}}.\EE
In addition, we used trainable cluster centers for the newly introduced classes.
These particular choices are also justified through ablation studies carried out in \autoref{subsec:ablation}.

\subsection{Feature exchange}
\label{subsec:featex}
\Ac{l3} embeddings \cite{arandjelovic2017look,arandjelovic2018objects,cramer2019look} are trained by using an audio and a video subnetwork, and predicting whether a video frame and an audio segment with a length of one second belong together or not.
In \cite{wilkinghoff2023using}, it was shown empirically that using these pre-trained embeddings does not lead to a better \ac{asd} performance than directly training an embedding model.
When comparing multiple pre-trained embeddings, self-supervised embeddings such as \ac{l3}-embeddings appear to outperform supervised embeddings.
This motivates to develop a similar \ac{ssl} approach for learning embeddings using only audio data.
\par
As the baseline system also consists of two sub-networks, both utilizing different input feature representations, a similar \ac{ssl} approach can be used, which we will call \acl{featex} (\acs{featex}).
Let ${e_1=(e^1_1,e^2_1)},{e_2=(e^1_2,e^2_2)\in\mathbb{R}^{2D}}$ with $D=128$ denote the concatenated embeddings of both sub-networks belonging to two random training samples $x_1$ and $x_2$, and let $y_1,y_2$ denote their corresponding categorical class labels.
Then, define a new embedding and its label by setting
\BE
e_\text{new}&=(e^1_1,e^2_2)\in\mathbb{R}^{2D}\\
y_\text{new}&=(\textbf{0},0.5\cdot y_1,0.5\cdot y_2)\in[0,1]^{3N_\text{classes}}
\EE
where $N_\text{classes}\in\mathbb{N}$ denotes the number of the original classes and $\textbf{0}=(0,...,0)\in[0,1]^{N_\text{classes}}$.
Hence, as for the \ac{statex} variant, the number of classes is tripled.
When applying \ac{featex}, the network also needs to learn whether the embeddings of the sub-networks belong together or not resulting in more information being captured.
Again, a combination with mixup, a probability of $50\%$ for applying \ac{featex} during training and trainable cluster centers for the newly introduced classes have been used throughout this work. 

\subsection{Combining supervised and self-supervised losses}
\label{subsec:combination}
In \cite{wilkinghoff2023design}, it was shown that not adapting randomly initialized cluster centers during training improves the resulting \ac{asd} performance.
Experimentally, we found that the \ac{ssl} approaches performed better with trainable cluster centers.
To ensure that only the cluster centers of the \ac{ssl} loss belonging to the original classes are non-trainable, we also used the regular supervised loss of the baseline system with non-trainable cluster centers as an equally weighted loss.
Since the classes introduced by the \ac{ssl} approaches are dividing the original classes into sub-classes, this can also be seen as a form of disentangled learning \cite{venkatesh2022improved}.
Furthermore, we propose to use a combination of mixup with the regular loss as well as \ac{statex} and \ac{featex} in a single loss function since all \ac{ssl} approaches are different.
Hence, the total loss $\mathcal{L}_\text{total}(x,y)$ of a sample $x$ with categorical label $y$ is given by
\BE\mathcal{L}_\text{total}(x,y)=\mathcal{L}(x,y)+\mathcal{L}(x_\text{new},y_\text{new})\EE
with $x_\text{new}$ and $y_\text{new}$ being defined by sequentially applying all the \ac{ssl} approaches as described in the previous sections and $\mathcal{L}$ denoting the categorical crossentropy.
As a result, the number of classes is increased multiplicatively by a factor of $3\cdot3=9$.
In the following, this is called the \emph{proposed approach}.

\section{Experimental results}

\subsection{Datasets}
\label{subsec:datasets}
For all experiments conducted in this work, the DCASE2022 \cite{dohi2022description} and the DCASE2023 \ac{asd} dataset \cite{dohi2023description} are used.
Both datasets are designed for semi-supervised \ac{asd} in machine condition monitoring and contain noisy recordings of machine sounds of various types taken from ToyAdmos2 \cite{harada2021toyadmos2} and MIMII-DG \cite{dohi2022mimii_dg}.
For training, only normal sounds and additional meta information such as the machine types and parameter settings of the machines, called attribute information, are available.
Furthermore, both datasets are designed for domain generalization and thus consist of data from a source domain with $1000$ training samples for each machine id and a target domain with only $10$ training samples that somehow differs by changing a parameter setting of the target machine or the background noise.
The task is to discriminate between normal and anomalous samples regardless of the domain a sample belongs.
\par
Both datasets are divided into a development and an evaluation set, each consisting of a training subset with only normal data samples and a test subset containing normal and anomalous samples.
The DCASE2022 \ac{asd} dataset consists of recordings belonging to seven different machine types, each with three different machine IDs contained in the development set and another three machine IDs contained in the evaluation set.
For the DCASE2023 \ac{asd} dataset, there are $14$ different machine types. The machine types belonging to the development and evaluation set are mutually exclusive, and there is only one machine ID for each machine type.
Hence, the classification task is much easier than for the DCASE2022 dataset and thus learning informative embeddings by solving an auxiliary classification task is much more difficult for the DCASE2023 dataset, which motivates to also utilize \ac{ssl} for training the \ac{asd} system.

\subsection{Comparison of \acs{ssl} approaches}
\label{sec:results}
As a first experiment, different \ac{ssl} approaches are compared to the baseline performance obtained by not using additional \ac{ssl} losses.
The results can be found in \autoref{tab:results} and the following observations can be made:
First, it can be seen that the proposed \ac{featex} loss performs significantly better than the \ac{statex} loss on both datasets.
Second, only using a loss based on \ac{statex} leads to slightly worse performance than the performance obtained with the baseline system while only using the \ac{featex} loss lead to slightly better performance than the baseline system for most dataset splits.
However, combining the regular loss with one of the \ac{ssl} losses improves performance over both individual losses, especially on the DCASE2023 dataset.
As stated before, the most likely reason is that, compared to the DCASE2022 dataset, there is only one specific machine for each machine type and thus the classification task is much easier resulting in less informative embeddings that are less sensitive to anomalies.
Hence, \ac{ssl} is required as a form of regularization to learn non-trivial mappings for each class resulting in more informative embeddings that enable the system to detect subtle deviations from normal data.
As a last observation, it can be seen that combining all \ac{ssl} approaches into a single loss, slightly improves the resulting performance for some dataset splits while decreasing the performance for other dataset splits.
Overall, the positive effects seem to be slightly greater than the negative ones but are only marginal.

\subsection{Ablation studies}
\label{subsec:ablation}
\begin{table}[t]
	\centering
	\caption{Harmonic means of AUCs and pAUCs taken over all machine types obtained on the DCASE2023 dataset by modifying design choices of the proposed approach. Arithmetic mean and standard deviation over five independent trials are shown.}
\begin{adjustbox}{max width=\columnwidth}
	\begin{tabular}{ll|ll|ll|ll}
		\toprule
		\multicolumn{2}{c}{}&\multicolumn{2}{c}{\acs{ssl} loss without class labels}&\multicolumn{2}{c}{non-trainable class centers}&\multicolumn{2}{c}{no \acs{tmn} and full \acs{statex}}\\
		split&domain&AUC&pAUC&AUC&pAUC&AUC&pAUC\\
		\midrule
        \midrule
		dev&source&$70.8\pm1.5$\%&$63.2\pm1.1$\%&$71.5\pm0.9$\%&$64.8\pm1.9$\%&$70.9\pm0.7$\%&$61.0\pm1.5$\%\\
		dev&target&$74.7\pm1.5$\%&$58.1\pm1.6$\%&$74.0\pm2.0$\%&$56.7\pm1.0$\%&$72.1\pm1.4$\%&$55.2\pm1.0$\%\\
		dev&mixed&$72.3\pm1.2$\%&$57.9\pm1.3$\%&$71.6\pm1.1$\%&$57.7\pm0.7$\%&$71.3\pm0.7$\%&$55.6\pm0.9$\%\\
		\midrule
		eval&source&$73.5\pm2.4$\%&$63.8\pm0.6$\%&$74.2\pm0.7$\%&$63.9\pm1.3$\%&$73.8\pm1.3$\%&$62.4\pm1.5$\%\\
		eval&target&$62.1\pm1.5$\%&$57.7\pm0.9$\%&$58.2\pm3.3$\%&$57.3\pm0.9$\%&$66.9\pm2.4$\%&$58.5\pm1.9$\%\\
		eval&mixed&$68.6\pm1.2$\%&$59.1\pm0.7$\%&$65.0\pm0.9$\%&$57.7\pm0.6$\%&$70.9\pm0.8$\%&$59.9\pm0.8$\%\\
        \midrule
		\bottomrule
	\end{tabular}
\end{adjustbox}
\label{tab:ablation}
\end{table}
To show that the design choices of the proposed approach actually optimize the \ac{asd} performance, three ablation studies have been conducted on the DCASE2023 dataset.
More concretely, it was verified whether 1) not using the class labels for the \ac{ssl} losses, 2) using non-trainable class centers for the \ac{ssl} losses or 3) not using \ac{tmn} but also applying \ac{statex} to the temporal axis improves the performance. 
When comparing the results, as shown in \autoref{tab:ablation}, to the original ones contained in \autoref{tab:results}, it can be seen that altering the proposed approach in any of the three ways degrades \ac{asd} performance, especially on the evaluation set.
This adds confidence to the design of the proposed \ac{ssl} approach.

\subsection{Comparison to other published systems}
\begin{figure}[t]
    \centering
    \begin{adjustbox}{max width=\columnwidth}
          \begin{tikzpicture}
\begin{axis}[
axis y line*=left,
axis x line*=bottom,
ybar=1,
ymin=50,
ymax=80,
xmin=0.9,
xmax=1.10,
area legend,
enlarge x limits=0.1,
point meta=explicit symbolic,
legend style={at={(1,0.5)},anchor=west,legend columns=1},
ylabel=official score in percent,
bar width=16.5pt,
height=5.8cm,
width=10cm,
xticklabels={},
xticklabel style={align=center},
yticklabel style={align=center},
xtick=\empty,
typeset ticklabels with strut,
xlabel near ticks,
ylabel near ticks,
every node near coord/.append style={/pgf/number format/precision=3,anchor=west,rotate=90, align=center},
nodes near coords,
ymajorgrids
]
\addplot[black,fill=red!85] coordinates {(1,70.93)[70.93\%]}; 
\addplot[black,fill=brown!50] coordinates {(1,66.97)[66.97\%]}; 
\addplot[black,fill=teal!85] coordinates {(1,66.39)[66.39\%]}; 
\addplot[black,fill=lightgray!50] coordinates {(1,65.40)[65.40\%]}; 
\addplot[black,fill=green!50] coordinates {(1,64.91)[64.91\%]}; 
\addplot[black,fill=darkgray!85] coordinates {(1,64.10)[64.10\%]}; 
\addplot[black,fill=magenta!50] coordinates {(1,63.64)[63.64\%]}; 
\addplot[black,fill=yellow!85] coordinates {(1,63.50)[63.50\%]}; 
\addplot[black,fill=olive!50] coordinates {(1,61.77)[61.77\%]}; 
\addplot[black,fill=blue!85] coordinates {(1,61.05)[61.05\%]}; 
\addplot[black,fill=pink!50] coordinates {(1,59.54)[59.54\%]}; 
\legend{
proposed system,
rank 1 \cite{junjie2023anomalous},
rank 2 \cite{lv2023unsupervised},
rank 3 \cite{jiang2023thuee},
rank 4 \cite{wilkinghoff2023fkie},
rank 5 \cite{yafei2023unsupervised},
rank 6 \cite{zhou2023attribute},
rank 7 \cite{tian2023first},
rank 8 \cite{wang2023first},
rank 9 \cite{harada2023first},
rank 10 \cite{JiaJun2023self},
}
\end{axis}
\end{tikzpicture}
    \end{adjustbox}
    \caption{Comparison between presented and ten top-performing systems of the DCASE Challenge 2023.}
    \label{fig:comp}
\end{figure}
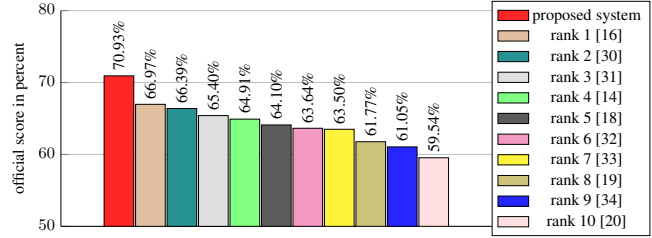
As a last experiment, the proposed system was compared to the ten top-performing systems submitted to the DCASE2023 Challenge.
To have a fair comparison, we used an ensemble obtained by re-training the system five times and taking the mean of all anomaly scores.
The results can be found in \autoref{fig:comp}.
It can be seen that our proposed system outperforms all other published systems by a significant margin and thus reaches a new state-of-the-art performance.
Note that the system ranked fourth \cite{wilkinghoff2023fkie} in the DCASE2023 Challenge is actually the same system as the baseline system of this work and the system ranked first \cite{junjie2023anomalous} is a modified version of this baseline system.

\section{Conclusion}
\label{sec:conclusion}
In this work, applying \ac{ssl} to \ac{asd} was investigated.
To this end, mixup and \ac{statex} were reviewed, and a novel \ac{ssl} approach for \ac{asd}, called \ac{featex} was proposed.
All three approaches were combined into a single loss function for training an outlier exposed \ac{asd} system.
In experiments conducted on the DCASE2022 and DCASE2023 \ac{asd} datasets, it was shown that \ac{featex} outperforms the existing \ac{ssl} approaches, and that applying \ac{ssl} to \ac{asd} is highly beneficial. 
As a result, a new state-of-the-art performance on the DCASE2023 \ac{asd} dataset was obtained outperforming all other published systems by a large margin.

\section{Acknowledgments}
The author would like to thank Fabian Fritz, Lukas Henneke and Frank Kurth for their valuable feedback.

\bibliographystyle{IEEEbib}
\bibliography{refs}

\end{sloppy}
\end{document}